\documentclass[12pt,article,aps,prd,tightenlines,indent,a4paper]{revtex4-2}
\usepackage[english]{babel}
\usepackage{blindtext}\blindmathtrue
\usepackage[dvipsnames]{xcolor}
\usepackage{xparse}
\usepackage{amsmath,epsf,amssymb,mathtools,latexsym,amsthm,setspace,array,pifont,hyperref,amsfonts,dsfont,cancel,braket,parskip}
\usepackage{slashed}
\usepackage{bbm}
\usepackage{graphicx}
\usepackage{float}
\def\be{\begin{equation}}
	\def\ee{\end{equation}}
	
\usepackage[title]{appendix}

\newcommand{\Rcal}{\mathcal{R}}
\newcommand{\Ocal}{\mathcal{O}}
\newcommand{\M}{\mathcal{M}}

\newcommand{\vmin}{\relax\ifmmode V^{(n)}_{\text{min}}\else $V^{(n)}_{\text{min}}$\fi}

\def\prd{\ref@{Phys.~Rev.~D}}        
\DeclareMathOperator{\Tr}{Tr}

\begin{document}

\title{Smoothed asymptotics: from number theory to QFT}

\author{Antonio Padilla}
\email{antonio.padilla@nottingham.ac.uk}
\author{Robert G. C. Smith}
\email{robert.smith@nottingham.ac.uk}
\affiliation{School of Physics and Astronomy, University of Nottingham, University Park, Nottingham NG7 2RD, United Kingdom}
\affiliation{The Nottingham Centre of Gravity, University of Nottingham, Nottingham NG7 2RD, UK}

\DeclareGraphicsRule{*}{mps}{*}{} 
\begin{abstract}
Inspired by the method of smoothed asymptotics developed by Terence Tao, we introduce a new ultra-violet regularisation scheme for loop integrals in quantum field theory which we call $\eta$ regularisation.  This reveals a connection between the elimination of divergences in divergent series of powers and the preservation of gauge invariance in the regularisation of loop integrals in quantum field theory. In particular, we note that a method for regularising the series of natural numbers so that it converges to minus one twelfth inspires a regularisation scheme for non-abelian gauge theories coupled to Dirac fermions that preserves the Ward identity for the vacuum polarisation tensor and other higher point functions. We also comment on a possible connection to Schwinger proper time integrals.
\end{abstract}

\maketitle

\section{Introduction}
The problem of infinity dates back to the sixth century BCE when Anaximander, a tutor of Pythagoras,  conjured up {\it apeiron}, an indefinite and limitless source from which everything is born and to which everything will return \cite{moore2018infinite}.  However, the ancient Greeks did not embrace the infinite with Aristotle accepting only the potential for infinity, rejecting it in actuality. Centuries later, Gauss, perhaps the greatest mathematician of the modern era, warned that ``the use of an infinite quantity  as a completed one ... is never permissible in mathematics. The infinite is only a {\it fa\c{c}on de parler}, where one is really speaking of limits where certain ratios come as close as one likes, while others are allowed to grow without restriction" \cite{waterhouse1979gauss}.

As physicists, how can we heed these warnings and at the same time make sense of divergent series and integrals that emerge unapologetically in the mathematics we use to describe fundamental physics? Perturbative quantum field theory (QFT) is well established as a microscopic theory of the fundamental interactions, enjoying predictive power and stunning experimental success \cite{ParticleDataGroup:2022pth}, and yet the presence of divergences is well documented.  Indeed, as Dyson noted \citep{Dyson52}, the perturbative expansion used in quantum electrodynamics  does not converge, even after renormalisation. Although this does not hinder its predictive power when the expansion parameter is small, it does raise concerns around the concepts upon which the theory is built.  Many notable advances have been made  in this regard using Borel summation, particularly under the heading of resurgence theory \citep{Dorigoni19,Dunne15qft,Marino14}. Here the usual perturbative expansion is replaced with a trans-series expansion, including non-perturbative instanton contributions. 

Even at finite order in perturbative QFT, we encounter divergences from the integration over momenta running through loops. These problems were first identified for quantum electrodynamics by Oppenheimer \cite{Oppenheimer:1930zz} and later solved using the method of renormalisation developed by Tomonoga \cite{Tomonaga:1946zz}, Schwinger \cite{Schwinger:1948iu} and Feynman \cite{Feynman:1948fi}, and extended to non-abelian gauge theories  by 't Hooft and Veltman \cite{tHooft:1972tcz}.  As for General Relativity, this works well as a perturbative quantum field theory at scales below the Planck scale, but cannot be extended to arbitrarily high energies as the theory is known to be perturbatively  non-renormalisable.  Said another way, for General Relativity, the number of counter-terms required to renormalise the divergences arising from loops is itself divergent!

 Adopting Aristotle's philosophy in rejecting actual infinity, at least in a natural context, one might hope that the correct theory of quantum gravity coupled to matter will be free of all such divergences. If this is the case, it may point towards a preferred way of regularising the   divergences in the low energy QFT.  String theory has certainly provided multiple insights in this regard, particularly in terms of a natural exponential damping of ultra-violet (UV) divergences, which can be seen, for instance, in the modular invariance of the worldsheet theory \citep{dhoker22,Polchinski98}.  Further, in the Gross-Mende regime \citep{GrossMende87,Gross:1987ar} where the high energy, fixed angle, behaviour of string scattering amplitudes is considered, the sum over all Riemann surfaces is dominated by a saddle point, and we recover the exponential damping from a stringy tower of states. With a view to understanding this from a simple particle perspective, Abel and others \citep{Abel19,Abel20} have constructed UV complete particle theories based on the Schwinger representation and  worldline formalism. This includes  particle theories with a tower of worldline internal degrees of freedom that mimic stringy behaviour in the UV, from Gross-Mende saddle points to modular invariance  \citep{Abel20}. Finally, string field theory has revealed how such exponential damping comes partly from adding stubs to loop diagrams \citep{Erbin21}, which can also be translated to the point particles of QFT \citep{Chiaffrino22}.

Mathematically rigorous and axiomatic formulations of perturbative QFT have progressed significantly in recent decades, aiding our understanding on the nature of UV divergences, with the view that all regularisation in QFT is really just an effort in obtaining sensible results for products of distributions \citep{Latorre94}. But, with this progress and insight, not much has been said fundamentally about regularisation besides it being a useful, if not altogether \textit{ad hoc}, tool.   This philosophy does not align with the hopes outlined in the previous paragraph and the quest to identify a preferred method of regularisation inherited from string theory, or some other consistent quantum theory of gravity. Where else can we take our inspiration? One possibility is through analytic number theory and the study of divergent series.  In particular, we  ask if there  are clever ways to regularise these series that eliminate the divergences altogether and if so, can we connect them, via QFT, to the softening of scattering amplitudes at high energies   in a fundamental microscopic theory of nature?

A particularly interesting approach to divergent series is the method of smoothed asymptotics, elegantly discussed by Tao \citep{Tao11}.  For monomial series, this makes use of the Euler-Maclaurin summation formula and as such is closely related to Ramanujan's methods \citep{Berndt85,Hardy73}. The importance of smoothed asymptotics in understanding divergent series is best illustrated with an example. Consider the infinite series of natural numbers, $\sum_{n=1}^\infty n$, which was famously assigned the seemingly absurd value of $-1/12$  by Ramanujan. The high school technique for evaluating infinite series is to take the limit of partial sums, which for the sum of natural numbers yields, $\sum_{n=1}^N n = \frac{1}{2} N (N+1)$. There is no sign of $-1/12$ in this expression, leading many to scoff at Ramanujan's claim. However, this is simply an artefact of the discontinuities that occur as we increase the cut-off $N$. To remedy this, we note that partial sums are really just infinite series where each term is weighted by a step function $\theta(n/N)$ that equates to unity for $n\leq N$ and vanishes for $n>N$. However,  suppose we replace the step function with another regulator function, $\eta(n/N)$, where $\eta(x)$ is a bounded smooth function with compact support on the non-negative real line and with $\eta(0)=1$ and $\eta(x) \to 0$ at large $x$. Tao shows that $\sum_{n=1}^\infty n \eta(n/N) = C_1[\eta] N^2-1/12+\mathcal{O}(1/N)$, where  $C_1[\eta]=\int_0^\infty dx \ x \eta(x)$ is the Mellin transform of the regulator function.  As we  will  explicitly show, the same method can also be extended to regulator functions that are Schwartz class.

Tao's results have several interesting features. The first concerns the divergence as $N \to \infty$. Unlike the partial sum, there is no linear divergence, while the quadratic divergence is dependent on the choice of regulator. This is in stark contrast to the finite term which is universal, returning Ramanujan's famous result.  These features extend to infinite series of polynomials and are reminiscent of a well known result regarding UV divergences in QFT. In particular, when we compute the one loop effective action for a theory cut-off at some scale, $\Lambda$, the couplings associated with power law divergences are not universal. This is in contrast to the logarithmic divergences where the couplings {\it are} universal, much like the finite terms in regularising divergent series of polynomials using smoothed asymptotics. This overlap suggests that one loop  divergences in QFT may be connected to divergent series. 

The power law divergences that emerge from regularising a series with smoothed asymptotics can also be eliminated with a suitable choice of regulator.  We call these enhanced regulators. For the series of natural numbers, an enhanced regulator is one for which the corresponding Mellin transform is vanishing $C_1[\eta]=0$. As we will show, such enhanced regulators are relatively easy to find and we present several algorithms for finding them.  A particularly elegant choice is the enhanced regulator $\eta(n/N)=e^{-\frac{n}{N}} \cos (n/N)$. When this regulator is used for the series of natural numbers the result {\it converges} to $-1/12$ and there is no divergence whatsoever!

To build these  ideas into QFT, we introduce the concept of $\eta$ regularisation. Here the integrand in (Euclidean) loop integrals is weighted by a regulator function $\eta(|k|/\Lambda)$, where $|k|$ is the norm of the loop momentum and $\Lambda$ is the cut-off.  It is convenient to work directly with  one fold irreducible loop integrals (ILIs) introduced by Wu \cite{Wu03} as the basic building block of all one particle  irreducible graphs.  When we implement $\eta$ regularisation we see, as expected, that power law divergences are regulator dependent while logarithmic divergences are universal.  As with divergent series and smoothed asymptotics, the power law divergences are multiplied by Mellin transforms of the regulator.  By choosing enhanced regulators, we can eliminate the power law divergences at will.  

However, the real question is whether or not the elimination of power law divergences by enhanced regulators has any deeper meaning from a QFT perspective.  Perhaps as one might have expected from dimensional regularisation, we find that the answer is yes.  In a series of papers \cite{Wu03,Wu04,Wu12,Wu13,Wu14,Wu17}, Wu and others have derived a set of consistency relations that are necessary in order for the regulator to preserve gauge invariance. Requiring that these hold for $\eta$ regularisation, we find that certain regulators must be enhanced.  In other words, regulators that allow the infinite series of natural numbers to converge towards $-1/12$ are intimately connected to the preservation of gauge invariance at one loop in a wide class of non-abelian gauge theories coupled to an arbitrary number of Dirac fermions.  This is the main result of our work.

The rest of the paper is organised as follows: in section \ref{sec:seriesrev}, we review several ideas from analytic number theory and the study of divergent series. We introduce a number of important concepts, including a brief review of the Euler-Maclaurin summation formula and Ramanujan summation in section \ref{sec:EM}. In section \ref{SectTaosmoothasymptotics}, we review Tao's work on smoothed asymptotics, extending the analysis to include regulators that are Schwartz and to series of general polynomials. We also introduce the concept of enhanced regulators that eliminate divergences altogether and present several algorithms for finding them, including one that is inspired by the Schwinger proper time formalism. In section \ref{SecDivIntegrals}, we switch gears to divergent integrals in QFT, drawing analogies between these and the results that emerged from regularising divergent series with smoothed asymptotics. We formally introduce $\eta$ regularisation in section \ref{sec:eta} and study the implications for irreducible loop integrals in section \ref{sec:ilis}.  In section \ref{VacuumPolarisation} we implement Wu's consistency conditions, demonstrating the role of enhanced regulators in preserving gauge invariance. We comment explicitly on the    connection to Schwinger proper time in section \ref{sec:SPT}. In section \ref{sec:conc}, we conclude.

\section{Divergent series: a physicist's review} \label{sec:seriesrev}

Let us begin with the following well known expressions for the finite sums of powers:
\begin{eqnarray}\label{Powersum_s1}
\sum_{n=1}^N n & =&  \frac{1}{2}N + \frac{1}{2}N^2,
\\
\sum_{n=1}^N n^2 & =& \frac{1}{6} N + \frac{1}{2}N^2 + \frac{1}{3}N^3, \label{Powersum_s2}
\\
\sum_{n=1}^N n^3  &=&  = \frac{1}{4}N^2 + \frac{1}{2}N^3 + \frac{1}{4}N^4. \label{Powersum_s3}
\end{eqnarray}
These relationships date back more than two millennia. The first of them, corresponding to a sum of natural numbers, can be traced to the Pythagorean school in the 6th century BCE  \citep{Peng02a, Sant17}. Archimedes of Syracuse (circa 287-212 BCE), considered the greatest mathematician of antiquity, discovered the second relationship for a sum of the squares. The sum of cubes can be found in the work of Nicomachus of Gerasa (circa 60 -120 CE),  along with the remarkable theorem that bears his name $$\sum \limits_{n=1}^{N} n^3 = \left(\sum \limits_{n=1}^{N} n\right)^2.$$

The expressions for the three finite sums  \eqref{Powersum_s1} to \eqref{Powersum_s3} are special cases of Faulhaber's formula \cite{Knuth_1993}
\begin{equation} \label{Faul}
\sum_{n=1}^N n^z=\frac{1}{z+1} \sum_{n=0}^z \binom{z+1}{n}  B_n N^{z-n+1},
\end{equation}
where $z$ is a positive integer and $B_n$ are the Bernoulli numbers (with $B_1= 1/2$), which are defined  by the following recursion relation
\be \label{recursion}
\sum_{n=0}^{z} \binom{z+1}{ n} B_{n}=z+1, \qquad B_0=1
\ee
or, equivalently, from  the exponential generating function
\be
\frac{t}{1-e^{-t}}=\sum_{n=0}^\infty B_n \frac{t^n}{n!}.
\ee
In the limit where we take $N \to \infty$, Faulhaber's formula will obviously break down.  This is because the infinite series $S(z)=\sum_{n=1}^\infty n^z$ is known to be divergent in the Cauchy sense  whenever $\Re(z) \geq -1$. According to Cauchy's definition,  an infinite series is said to be convergent if the sequence of partial sums converges to some finite limit; otherwise, the series is divergent. If the series is not convergent in the Cauchy sense, then it will generally be considered as one of two types of divergent series \citep{Hardy73,Graham91ConcreteMaths}: (i) a series that grows in absolute value without limit,  or (ii) a series that is bounded but whose sequence of partial sums does not approximate any specific value. 

At high school we are taught to think of an infinite series $\sum_{n=1}^\infty a_n$ as the limit of its partial sums, $\lim_{N\to\infty} \sum_{n=1}^N a_n$.  Partial sums are favoured because  they allow us to  perform  standard arithmetic without issue. Of course, the method works well for convergent series, less so for divergent series.  With the latter the task is to develop a summation method that shares the most important properties of partial sums but allows the result to be generalised to infinity without giving a divergent answer.  Alternative summation methods include Cesaro summation, where we compute the sequence of partial sums $s_N=\sum_{n=1}^N a_n$ and find the limiting value of their average $\lim_{N \to \infty} \frac{1}{N}\sum_{n=1}^N s_n$; and Abel summation, where we compute $\lim_{t \to 1^-} \sum_{n=1}^\infty a_n t^n$. Hardy has argued that any new summation method should satisfy three properties: regularity, linearity and stability \citep{Hardy73}. Regularity states that a summation method yields the known results for convergent series obtained using partial sums. For linearity, we  require that $\sum_{n = 1 }^\infty {\lambda a_n+\mu b_n=}\lambda \sum_{n = 1 }^\infty a_n+\mu \sum_{n = 1 }^\infty b_n$.  The third property of stability, namely  $\sum_{n = 1 }^\infty a_n =a_1+\sum_{n = 1 }^\infty a_{n+1}$,  is considered less crucial and known to fail in some important cases, including Ramanujan summation \citep{Candelpergher17}.  

 The real question is whether any meaning can  be extracted from the finite results obtained by these alternative methods. This is more than just a mathematical curiosity. In Physics, the mathematical description of  physical phenomena is often given in terms of divergent series and/or integrals. Although the formal treatment of divergent series remains an open question, a  commitment to truncated partial sums ignores the fact that divergences can lead to inconsistent results in physical situations. Of course, such divergences may just be a reflection of a breakdown in the relevant mathematical description. However, in some cases, physically consistent results can be obtained by using alternative summation methods that yield finite  as opposed to infinite answers. Indeed, the infinite series of natural numbers appears in computations of the Casimir force \cite{Schwartz14} and in the critical dimension of bosonic string theory \cite{Polchinski98}, where it is necessarily interpreted as the seemingly absurd formula $\sum_{n=1}^\infty n = -1/12$ \citep{Hardy73,Berndt85}. 
 
 Although  $\sum_{n=1}^\infty n = -1/12$ is often associated with Ramanujan \cite{Berndt85}, the result is best derived using zeta function regularisation.  When $\Re(z)>1$, the Riemann zeta function is defined by the convergent series $\zeta(z)=\sum_{n=1}^\infty n^{-z}$. Of course, when $\Re(z) \leq 1$ the series is no longer convergent. However, we can analytically continue $\zeta(z)$ into $\Re(z)<1$ using the following formula
 $$
 \zeta(z)=2^z \pi^{z-1} \sin\left(\frac{\pi z}{2} \right) \Gamma(1-z) \zeta(1-z).
 $$
 We may now identify the series given above,  $S(z)=\sum_{n=1}^\infty n^{z}$ with $\zeta(-z)$, even when it is divergent, $\Re(z)\geq -1$.  In particular,  we obtain $S(1)=\sum_{n=1}^\infty n \overset{\zeta}{=}\zeta(-1)=-1/12$.  More generally, to sum a divergent series $\sum_{n=1}^\infty a_n$ using zeta function regularisation, we identify a convergent series $\sum_{n=1}^\infty a_n^{-z}$ for $z$ in some complex domain, then analytically continue the result to $z=-1$.

The majority of issues with divergent series come with the transition from partial sums to infinity, as notably exposed by Ramanujan \citep{Berndt85,Hardy73}. Although it does not appear that he  fully understood the validity of his  asymptotic expansions, he was able to extract some remarkable results. As we will see in a moment,  Ramanujan developed a study of divergent series based upon the Euler-Maclaurin summation formula, often employing it in creative ways.  The Euler-Maclaurin formula also plays an important role in understanding the results of smoothed asymptotic expansions, an intuitive approach to the regularisation of divergent series beautifully elucidated by Tao \citep{Tao11}. Indeed, using Tao's methodology we will begin to understand how a divergent series of positive numbers could ever be identified with a finite negative number in a meaningful way. 

\subsection{Ramanujan and the Euler-Maclaurin summation formula} \label{sec:EM}
In the early 1730s, Euler solved one of the most intriguing mathematical puzzles of the time: the Basel problem  \citep{Euler08Reciprocals}.  First formulated by Leibniz and the Bernoulli brothers \citep{Peng02a}, the Basel problem concerns the infinite series whose terms are the reciprocal squares of the natural numbers. Using calculus to relate the discrete sum of an arbitrary function $\sum_n \limits f(n)$ to an integral of the form $\int  dx  f(x)$, Euler was able to prove that $\sum_{n=1}^\infty \frac{1}{n^2}=\frac{\pi^2}{6}$. For a historical review of the development of some of Euler's key ideas, see \citep{Peng07,Ferraro98}.

From elementary calculus, we know that a sum and an integral provide first approximations to one other. In particular,  the sum can be interpreted as the total area of rectangles forming a step graph along some curve, while the integral, evaluated between end points, can  be interpreted as the area under the same curve. More formally, the Euler-Maclaurin summation formula provides an estimation of the sum $\sum \limits_{n=a}^b f(n)$ in terms of the integral $\int_a^b \ dx \ f(x)$ and the derivatives of the function $f(x)$. As an estimate of how much the trapezoid rule fails, an error term is also included given by an integral involving Bernoulli polynomials. The formula can be written as \citep{Candelpergher17}
\begin{equation}\label{EulerMaclaurinfinal}
\sum_{n=a}^b f(n) = \int^b_a dx f(x) + \frac{f(b) + f(a)}{2}+ \sum \limits_{k=2}^{m} \frac{B_{k}}{k!} (f^{(k-1)}(b)-f^{(k-1)} (a))  + R_m,
\end{equation}
where   $a,b$ are integers such that $a \leq b$ and the integer $m \geq 2$.  We denote the $k$th derivative $f^{(k)}(x)=\frac{d^k}{dx^k}f$.  The sum on the right-hand side contains the non-vanishing Bernoulli numbers $B_{2n}$ (recall that $B_{2n+1}=0$ for $n \geq 1$). As these numbers grow asymptotically for large $k$ as $B_{2n} \sim (-1)^{n+1} 4 n^{2n} (\pi e)^{-2n} \sqrt{\pi n }$, this sum often diverges as $m \to \infty$, and is best treated as an asymptotic series. The corresponding error or remainder  is given by 
\be
R_m = (-1)^{m+1} \int^b_a dx \ \frac{b_{m}(x)}{m!} f^{(m)}(x),
\ee
where $b_m(x)$ is the periodic function $B_m(x-\lfloor x \rfloor)$ given in terms of the Bernoulli polynomials. The latter are defined by the generating function
\be
\frac{te^{xt} }{e^{t}-1}=\sum_{n=0}^\infty B_n(x) \frac{t^n}{n!}. 
\ee
 As emphasized in \cite{Tao11}, the level of approximation offered by \eqref{EulerMaclaurinfinal} is heavily dependent on the asymptotic behaviour of $R_m$. It is often the case that the expansion remains valid even after taking the limits $a \rightarrow -\infty$ and/or $b \rightarrow \infty$. Generally, the integral on the right-hand side can be evaluated in closed form in terms of elementary functions, even though the sum on the left-hand side cannot. In the best case  scenario, all the terms in the asymptotic series can be expressed in terms of elementary functions. 
 
The idea behind Ramanujan's summation method is to use the Euler-Maclaurin formula in the following asymptotic form
\be \label{EMas}
\sum_{n=1}^N f(n) \sim \int^N_a dx f(x) + \frac{f(N)}{2}+ \sum \limits_{k=2}^{\infty}  \frac{B_{k}}{k!} f^{(k-1)}(N)+C(f, a),
\ee
where $C(f, a)$  is the so-called ``constant of  the series".  This will depend on the choice of  lower limit on the integral, first 
noted by Hardy \cite{Hardy73}. We will return to this ambiguity in a moment. This constant is identified with the corresponding  infinite series in the limit where $N \to \infty$ such that
\be
\sum_{n=1}^\infty f(n) \overset{(\mathcal{R}, a)}{=} C(f, a).
\ee
The $\Rcal$ corresponds  to Ramanujan, whereas the $a$ reflects the ambiguity in choosing the limits on the integral. To recover $C(f, a)$, we first  define a sequence of constants $C_m(f,a)$ via the Euler-Maclaurin formula. In particular, we write
\be
\sum_{n=1}^N f(n) = \int^N_a dx f(x) + \frac{f(N)}{2}+\sum \limits_{k=2}^m   \frac{B_{k}}{k!} f^{(2k-1)}(N)+(-1)^m \int^\infty_N dx \ \frac{b_{m}(x)}{(2m)!} f^{(m)}(x)+C_m(f, a),
\ee
so that 
\be
C_m(f, a)=-\int^1_a dx f(x) + \frac{f(1)}{2}-\sum \limits_{k=2}^m    \frac{B_{k}}{k!} f^{(k-1)}(1)+(-1)^{m+1}\int^\infty_1 dx \ \frac{b_{m}(x)}{m!} f^{(m)}(x).
\ee
Assuming that $f \in \mathcal{C}^\infty$ and the integral above is  convergent for sufficiently large values of $m$,  one is able to show that $C_m(f, a)$ becomes independent of $m$ at large $m$, settling down to  $C(f, a)$. For $f(n)=n^z$, where $z$ is a positive integer,  it is sufficient to take any $m \geq z+1$ and evaluate the constant of the series, giving 
\be \label{Intnz}
\sum_{n=1}^\infty n^z \overset{(\mathcal{R}, a)}{=} \frac{1}{z+1} (a^{z+1}-1)+\frac12 - \frac{1}{z+1}\sum \limits_{k=2}^{z+1}   \binom{z+1}{ k} B_{k}=\frac{a^{z+1}}{z+1}-\frac{B_{z+1}}{z+1}
\ee
where we have used the recursion relation \eqref{recursion} for the Bernoulli numbers. Given that $\zeta(-z)=-\frac{B_{z+1}}{z+1}$ for positive integers, $z$,  we see the connection of this result to zeta function regularisation in the limit where  $a=0$.

To better understand the role played by the arbitrary real number $a$ in general, we first note that
\be
C(f, b)-C(f, a)=\int_a^b dx f(x).
\ee
Furthermore, when the series $\sum_{n=1}^\infty f(n)$ is convergent, it is easy to see from the asymptotic formula \eqref{EMas} that \cite{Candelpergher17}
\be
C(f, 1)=\sum_{n=1}^\infty f(n) - \int^\infty_1 dx f(x).
\ee
It immediately follows that $\sum_{n=1}^\infty f(n) =C(f, \infty)$, demonstrating the fact we should choose $a=\infty$ for the case of convergent series. For a divergent series with  $f(n)=\sum_{z=0}^s c_z n^z$ a polynomial of degree  $s$,  it is clear that we must take $a=0$ in order to match the result obtained via analytic continuation.  When we develop the method of smooth asymptotics advocated by Tao \cite{Tao11} in the next section, we will see how $C(f, 0)$ also picks out the finite term that is independent of the choice of cut-off.

\subsection{Smoothed asymptotics}\label{SectTaosmoothasymptotics}
When we use partial sums to sum an infinite series $\sum_{n=1}^\infty a_n$, we truncate the series at some finite value $N$ and then compute the sum $\sum_{n=1}^N a_n$ before taking $N \to \infty$. We can think of the partial sum as modifying the infinite series with a step function  
\be
\theta(x)=\begin{cases} 
1 & x \leq 1  \\
0  &x>1
\end{cases}
\ee
such that
\be
\sum_{n=1}^\infty a_n \to \sum_{n=1}^\infty a_n \theta\left(\frac{n}{N}\right)=\sum_{n=1}^N a_n.
\ee
In the case of a convergent series, the partial sum  tends towards a unique finite value  as we take larger and larger values of $N$.  For a divergent series, we have seen how there exist alternative summation that yield finite answers. Unfortunately, there is no obvious trace of those finite values anywhere in the partial sum.  This is readily seen in Faulhaber's formula \eqref{Faul} where we see no evidence for the finite value of $S(z)=\sum_{n=1}^\infty n^z$ obtained via analytic continuation. In \cite{Tao11}, Tao argues that this is an artefact of the jump discontinuity in $\theta(x)$ and can be remedied by generalising $\theta(x)$ to a smooth regulator function, as opposed to a sharp cut-off. 

To this end, we regularise the infinite series with a smooth function, $\eta(x)$, such that
\be
\sum_{n=1}^\infty a_n \to \sum_{n=1}^\infty a_n \eta\left(\frac{n}{N}\right). 
\ee

We assume that  $\eta(x)$ is  a smooth bounded function, defined for the non-negative real numbers, with $\eta(0) = 1$ and $\eta(x) \to 0$ at large $x$.  Indeed, Tao assumes that $\eta(x)$ is a bump function - a smooth function with compact support, vanishing whenever $x \notin [0,1]$.  Here we will relax this condition a little, and allow $\eta(x)$ to be a Schwartz function. That means that $\eta(x)$ and all its derivatives are rapidly decreasing at large $x$,  going to zero at infinity faster than $x^{\alpha}$ for all $\alpha < 0$.

For series that were already absolutely convergent, the smoothing by $\eta(x)$  does not affect the asymptotic value \cite{Tao11}. However, for divergent series, smoothing can significantly improve the convergence properties.  For example, we can  easily sum Grandi's series $\sum_{n=1}^\infty (-1)^{n-1}$ using a regulator function $\eta(x)=e^{-x}$, so that the corresponding series converges towards one half. 

Let us consider the impact of smoothing on a divergent series $\sum_{n=1}^\infty f(n)$ where $f(n)=\sum_{z=0}^s c_z n^z$ is a polynomial of degree  $s$. From the Euler-Maclaurin formula \eqref{EulerMaclaurinfinal} applied to the function $g_N(x)=f(x)\eta(x/N)$ with $a=0$ and $b=\infty$, we readily obtain the following expression
\begin{equation}\label{EM}
\sum_{n=1}^\infty g_N(n)= \int^\infty_0 dx \ g_N(x) - \frac{g_N(0)}{2}- \sum \limits_{k=2}^{m}  \frac{B_{k}}{k!} g_N^{(k-1)} (0)+ R_m,
\end{equation}
where we have used the fact that $g_N(x)$ and all of its derivatives vanish at infinity and the remainder term is given by 
\be
R_m= (-1)^{m+1} \int^\infty_0 dx \ \frac{b_{m}(x)}{m!} g_N^{(m)}(x).
\ee
As $g_N^{(k)}(x)=f^{(k)}(x)\eta(x/N)+\mathcal{O}(1/N)$, and given the polynomial form for $f(n)=\sum_{z=0}^s c_z n^z$, we choose $m \geq s+2$ and find that
\begin{equation}\label{EM2}
\sum_{n=1}^\infty f(n)\eta\left(\frac{n}{N}\right)= \sum_{z=0}^s c_z \left[C_{z}[\eta]   N^{z+1} +\zeta(-z)\right]+\mathcal{O}(1/N),
\end{equation}
where  $C_{z}[\eta] =\int_0^\infty dx  \ x^z \eta(x)$ is the Mellin transform of the regulator function. Here we have used the fact that $\zeta(-z)=-\frac{B_{z+1}}{z+1}$ for natural numbers $z$. The appearance of the zeta function follows in a very similar way to its appearance in equation \eqref{Intnz}. In choosing $m \geq s+2$, we guarantee that the remainder term $R_m =\mathcal{O}(1/N)$.  This is not immediately obvious. In particular we find that 
\begin{eqnarray}
R_{m} &=& (-1)^{m+1}  \sum_{z=0}^s c_z  \int^\infty_0 dx \ \frac{b_{m}(x)}{m!} \frac{d^m}{d x^m} \left[  x^z   \eta\left(\frac{x}{N}\right)\right] \\
&=& (-1)^{m+1}  \sum_{z=0}^s c_z  N^{z+1-m} \int^\infty_0 dy \ \frac{b_{m}(Ny)}{m!} \frac{d^m}{d y^m} (  y^z   \eta\left(y\right)).
\end{eqnarray}
Now since $\eta$ is Schwartz and $b_m$ is  bounded, it follows that the integrand $\frac{b_{m}(Ny)}{m!} \frac{d^m}{d y^m} (  y^z   \eta\left(y\right))$ is also Schwartz, and so the corresponding integral is bounded\footnote{If $F(x)$ is Schwartz then for each $n\geq 0$,  $|x^n F(x)|<Q_{n}$ for finite constants $Q_n$. Consider the corresponding integral $\int_0^\infty dx F(x)= \int_0^1dx F(x) +\int_1^\infty dx F(x)$. $F(x)$ is smooth so the first integral running over a finite range, $ \int_0^1dx F(x) $,  is bounded. As for the second integral, note that $\left| \int_1^\infty dx F(x)\right| \leq \int_1^\infty dx |F(x)| < \int_1^\infty dx Q_n/x^n=Q_n/(n-1)$.  Taking $n \geq 2$ we immediately see that  $\int_1^\infty dx F(x)$ is bounded. It follows that  $\int_0^\infty dx F(x)$ is bounded.}. Since $m \geq s+2$, and $s$ is finite,  we immediately see that $R_m =\mathcal{O}(1/N)$.

For monomials $f(n)=n^z$, there is a single divergence as $N\to \infty$ in the corresponding expression for the regularised series \eqref{EM2}, and there is a unique finite piece given by $\zeta(-z)$. This coincides with the result obtained using Ramanujan or analytic continuation.  The connection to Ramanujan summation (with $a=0$) is understood via the following identity derived directly from \eqref{EM2}
\be
\sum_{n=1}^\infty n^z\overset{(\Rcal, 0)}{=}C(f, 0)=\lim_{N \to \infty}\left[\sum_{n=1}^\infty n^z \eta\left(\frac{n}{N}\right)-\int_{0}^\infty dx  x^z \eta\left(\frac{x}{N}\right)\right]=\zeta(-z).
\ee
Using the method of smoothed asymptotics, we see how Ramanujan performed a delicate cancellation between infinities that can be rigorously understood. The procedure is reminiscent of renormalsation in perturbative QFT. 

As we will explore in more detail in the next section, another feature of equation \eqref{EM2} reminiscent of QFT is the regulator dependence in the power law divergences. 
%
%
The regulator dependence in the divergences of \eqref{EM2} raises the possibility that there are families of {\it enhanced} regulators for which the power law divergences vanish altogether, just as they do for dimensional regularisation.

This is indeed the case: an enhanced regulator is one for which the Mellin transform $C_{z, \eta} =\int_0^\infty dx \  x^z \eta(x)$  vanishes for integer values of $z \geq 0$. More precisely, for all natural numbers $z$, we define an {\it enhanced regulator of order $z$} to be a regulator function $\eta_{[z]}(x)$ for which 
$
\int_0^\infty dx \ x^z \eta_{[z]}(x)=0
$. We immediately see from \eqref{EM2} that if we use an enhanced regulator of order $z$ to  regularise the monomial series $S(z)=\sum_{n=0}^\infty n^z$,  there are no divergences whatsoever. Indeed, the regularised series converges to the finite  value $ \zeta(-z)$ as $N\to \infty$, without any need to renormalise.

An elegant example of an enhanced regulator of order $z$ is  given by the function 
\be \label{etaz}
\eta_{[z]}(x)=e^{-x\cot \left(\frac{\frac{\pi}{2}-\theta}{z+1}\right)}\frac{\cos(x+\theta)}{\cos\theta}, \ee
where $0< \theta<\frac{\pi}{2}$ and $z$ is any natural number. One can readily check that  we have $\int_0^\infty dx  \ x^z \eta_{[z]}(x)=0$ and so the smoothed series $\sum_{n=1}^\infty n^z \eta_{[z]}\left(\frac{n}{N}\right)$ converges to $\zeta(-z)$.   We can actually take $\theta$ to be zero, except in the case of $z=0$ where non-vanishing $\theta$ is required in order to ensure that the regulator remains Schwartz. Note that for  $\theta=0$ and $z=1$ we recover the rather beautiful enhanced regulator of order one, $\eta_{[1]}(x)=e^{-x} \cos x$, from which we infer that\footnote{To our best knowledge this remarkable regulator does not formally appear in the literature. However, in the process of this research, we found this formula appears on \href{https://en.wikiversity.org/wiki/MATLAB/Divergent_series_investigations}{Wiki} without citation or context. It is also the topic of discussion on this mathematics.stackexchange \href{https://math.stackexchange.com/questions/1327812/limit-approach-to-finding-1234-ldots}{post}, in which the author notices the nature of the function being Schwartz and also relates it to Tao's method of smooth summation.}
$$\lim_{N \to \infty} \sum_n n e^{-\frac{n}{N}} \cos \left(\frac{n}{N} \right)=-\frac{1}{12}.$$ 

 Since $\int_0^\infty dx \  x^z e^{-x} \cos x=2^{-\frac12(1+z)}z! \cos \left(\frac{\pi}{4} (z+1)\right)$ it just so happens that $e^{-x} \cos x$ is also an enhanced regulator of order $z=4m+1$ for any natural number $m$. Likewise, it turns out that $\eta_{[z]}(x)=e^{-x\cot \left(\frac{\pi}{2(z+1)}\right)}\cos x $  is also enhanced regulator of order $z'=z+2(z+1)m$ for any natural number $m$ and $z \geq 1$.

More generally, we can employ a few simple algorithms for finding enhanced regulators of any given order. The first assumes that we already know an enhanced regulator of order zero, $\eta_{[0]}(x)$, with vanishing integral $\int_0^\infty dx \eta_{[0]}(x)=0$. By changing variables from $x \to x^{z+1}$, we see that $\eta_{[z]}(x)=\eta_{[0]}(x^{z+1})$ is an enhanced regulator of order $z$.  
The second algorithm requires no previous knowledge of enhanced regulators: take any Schwartz function $\chi(x)$ defined on the positive real axis with the property that $\chi(0)=0$  and  $\chi^{(z+1)}(0)=1$, then $\eta_{[z]}(x)=\chi^{(z+1)}(x)$ is an enhanced regulator of order $z$. To see this, we first note that since $\chi(x)$  is a Schwartz function, the same is trivially true of $\chi^{(z+1)}(x)$. Furthermore, since $\chi^{(z+1)}(0)=1$, it is clear that $\chi^{(z+1)}(x)$ has all the properties of a regulator as defined at the beginning of this section. To see that it is an {\it enhanced} regulator, we compute the relevant Mellin transform
\be
\int_0^\infty  dx \ x^z\chi^{(z+1)}(x)=\left[ \sum_{k=0}^{z} (-1)^k \frac{z!}{(z-k)!} x^{z-k} \chi^{(z-k)}(x)\right]^\infty_0=0
\ee
verifying the defining condition for an enhanced regulator of order $z$. 

A third and final algorithm connects enhanced regulators to Schwinger's proper time formalism \cite{SchwingerPT}. Indeed,  for any natural number $n$, consider a regulator of the form
\be
\lambda_n(x)=\frac{1}{(n-1)!}\int_0^\infty \frac{du}{u}  \rho(u)  (ux^2)^{n} e^{-ux^2} \label{lambdan},
\ee
where the bounded function $\rho(u) \to 0$ faster than any power as $u\to 0$ and $\rho(u) \to 1$ as $u \to \infty$. The properties of $\rho(u)$ guarantee  that this function is Schwartz with $\lambda_n(x) \to 1$ as $x \to 0^+$, as required for a regulator.  After changing variables from $x$ to $ux^2$,  it is not difficult to show that the Mellin transform is given by
\be \label{Mellinlam}
\int_0^\infty dx \ x^z \lambda_n(x)= \frac{\Gamma\left( n+\frac{z+1}{2}\right)}{2(n-1)!}\int_0^\infty du \ \rho(u)u^{-\left(\frac{z+3}{2}\right)}.
\ee
For any natural numbers $n\neq m$, we can now construct an enhanced regulator of order $z$ as follows
\be \label{enhlam}
\eta_{[z]}(x)=\frac{\frac{(n-1)!}{\Gamma\left( n+\frac{z+1}{2}\right)}\lambda_n(x)-\frac{(m-1)!}{\Gamma\left( m+\frac{z+1}{2}\right)}\lambda_m(x)}{\frac{(n-1)!}{\Gamma\left( n+\frac{z+1}{2}\right)}-\frac{(m-1)!}{\Gamma\left( m+\frac{z+1}{2}\right)}}.
\ee 
This is clearly normalised so that it is classed as a regulator. To see that it is enhanced we simply check that the corresponding Mellin transform vanishes $\int dx \  x^z\eta_{[z]}(x)=0$ using \eqref{Mellinlam}.

Although the enhanced regulators described above work well for summing specific monomial series, they are less suited to eliminating divergences for generic polynomial series. For example, the regulators given in equation \eqref{etaz} will yield convergent series for $\sum_n n^z \eta_{[z]}(n/N)$ but not for, say,  $\sum_n (n^{z-1}+n^z)\eta_{[z]}(n/N)$. To remedy these problems for all natural numbers $ r \leq s$, we introduce the notion of an enhanced regulator  of order $[r, s]$ defined to be a regulator function $\eta_{[r, s]}(x)$ for which
$
\int_0^\infty  dx  \ x^z \eta_{[r, s]}(x)=0
$
for {\it all} natural numbers $z  \in [r,  s]$. It now follows from \eqref{EM2} that if we use an enhanced regulator of order $[0,s]$ to  regularise the series over polynomials of degree $s$,  there are no divergences whatsoever. The regularised series for the polynomial  $f(n)=\sum_{z=0}^s c_z n^z$ converges to the finite  value $\sum_{z=0}^s c_z \zeta(-z)$ as $N\to \infty$ without any need to renormalise. For the example given above, we find that $\lim_{N \to \infty} \sum_n (n^{z-1}+n^z)\eta_{[0, z]}(n/N)=\zeta(1-z)+\zeta(-z)$ for $z\geq 1$.

We can generalise the second algorithm  described above to find enhanced regulators of order $[r, s]$: take any Schwartz function $\chi(x)$ defined on the positive real axis with the property that $\chi^{(k)}(0)=0$ for all natural numbers $k \in [0,s-r]$ and with $\chi^{(s+1)}(0)=1$. It follows that $\eta_{[r,s]}(x)=\chi^{(s+1)}(x)$ is an enhanced regulators of order $[r,s]$.  The argument for $\eta_{[r,s]}(x)$ being a regulator is the same as for $\eta_{[z]}(x)$. To see that it is enhanced at order $[r,s]$, we compute the Mellin transforms for all natural numbers $z\in [r, s]$ and see that they all vanish
\be
\int_0^\infty dx  \ x^z\chi^{(s+1)}(x)=\left[ \sum_{k=0}^{z} (-1)^k \frac{z!}{(z-k)!} x^{z-k} \chi^{(s-k)}(x)\right]^\infty_0=0.
\ee
As an example, we consider the case where $\chi(x)=e^{-x} \frac{x^{s+1}}{(s+1)!}$ which generates an enhanced regulator of order $[0,s]$ given by
\be
\eta_{[0, s]}(x)=e^{-x} \sum_{k=0}^{s+1} \binom{s+1}{k} \frac{(-x)^k}{k!}.
\ee
The existence of enhanced regulators challenges the idea that divergent series of polynomials must be renormalised in order to recover their corresponding finite values. Indeed, they  explicitly demonstrate  a rigorous method of taking the series towards infinity without encountering any divergence. Of course, there is a magic of sort: the enhanced regulators must always change sign and in just the right way in order for large positive contributions to be cancelled by large negative ones.  It is here that we are reminded of the words of Euler who said ``the quantities greater than infinity are also smaller than nothing and the quantities smaller than infinity also correspond to the quantities greater than nothing"\citep{Euler18DeSeriebusDivergentibus}.

\section{Divergent integrals in quantum field theory}\label{SecDivIntegrals}
Divergent integrals are encountered with unsettling regularity in interacting QFTs. As revealed by Tomonoga \cite{Tomonaga:1946zz}, Schwinger \cite{Schwinger:1948iu} and Feynman \cite{Feynman:1948fi} in the context of quantum electrodynamics and by 't Hooft and Veltman for non-abelian gauge theories \cite{tHooft:1972tcz},  for renormalisable theories, these divergences can be carefully regularised and absorbed into a renormalisation of a finite number of couplings.  The origin of these divergences is understood to be a consequence of pushing the corresponding theory too hard, and assuming it to apply even at the shortest distance scales, or equivalently, at the highest energies.  Thanks to the Applequist-Carazzone decoupling theorem, the effects of very heavy particles on low energy scattering processes are hidden inside a renormalisaton of the low energy couplings \cite{Appelquist:1974tg}.  Said another way, even though our QFTs encounter divergences that betray limitations on their validity at all scales, they can still be made to work well as effective field theories at low energies.

It is interesting to note parallels between divergences in effective field theory and the characteristic behaviour we see whenever we regularise a divergent series with a smooth cut-off.  Indeed, it is well known that power law divergences in QFT are regulator dependent in stark contrast to the universal prediction obtained from the logarithmic divergences. Although not universal, power law divergences do play a role in the correct matching  to UV physics in the Wilsonian EFT. They can  also indicate an unwelcome sensitivity to the details of the UV completion, leading to problems with naturalness in high energy physics and cosmology \cite{Giudice}.

In order to develop the analogy with our divergent series, it is instructive to examine  power law and logarithmic divergences in QFT in a little more detail. Closely following the discussion presented in \cite{deRham:2014wfa}, we consider the Wilson action $S_\Lambda[\varphi]$ for modes lighter than some scale $\Lambda$ defined according to the path integral \cite{Burgess:2020tbq}
\be
e^{iS_\Lambda[\varphi]}=\int_{k>\Lambda} D [\varphi]e^{iS[\varphi]}.
\ee
This  generically contains power law and logarithmic divergences, and can be written as
\be
S_\Lambda[\varphi]=\int d^4 x \left[ \sum_i \Lambda^{4-d_i} g_i (\Lambda) \Ocal_i(x)+\ln(\Lambda/\mu)\sum_a g_a (\Lambda) \Ocal_a(x)+\ldots \right],
\ee
where $\Ocal_{i}$ are relevant operators of dimension  $d_i=0, 2$ and $\Ocal_{a}$ are marginal operators of dimension $4$.  The ellipsis denote finite terms which may also depend on $\mu$, an arbitrarily chosen  mass scale. Now consider the Wilson action, $S_\Lambda'[\varphi]$, defined at some lower scale, $\Lambda'<\Lambda$, and given by
\be
e^{iS_\Lambda'[\varphi]}=\int_{\Lambda'<k<\Lambda} D [\varphi]e^{iS_\Lambda[\varphi]}.
\ee
This is obviously independent of the choice of $\Lambda$, but highly dependent on $\Lambda'$ and correspondingly takes the form
\be
S_{\Lambda'}[\varphi]=\int d^4 x \left[ \sum_i \Lambda'^{4-d_i} g_i (\Lambda') \Ocal_i(x)+\ln(\Lambda'/\mu)\sum_a g_a (\Lambda') \Ocal_a(x)+\ldots \right].
\ee
It follows that $S_\Lambda'=S_\Lambda+\Delta \Gamma$ where $\Delta \Gamma$ is the contribution from integrating out intermediate modes $\Lambda' \leq k \leq \Lambda$. Focussing on the relevant and marginal operators, we must have
\be
\Delta \Gamma = \int d^4 x \left[ \sum_i c_i(\Lambda', \Lambda) \Ocal_i(x)+\sum_a c_a(\Lambda', \Lambda) \Ocal_a(x)+\ldots \right],
\ee
where the contribution from the intermediate modes $c_i(\Lambda', \Lambda)=\Lambda'^{4-d_i} g_i (\Lambda') -\Lambda^{4-d_i} g_i (\Lambda)$ completely cancels the power law divergences in passing from one Wilson action to another.  This is just Wilsonian renormalisation in action -  the coefficients of power-like divergences in the momentum cut-off get cancelled in the regularisation procedure (see \cite{Branchina} for elegant discussions on this point). In contrast, the cancellation of the $\Lambda$ dependence for the  logarithmic couplings yields $c_a (\Lambda', \Lambda)= \ln(\Lambda'/\mu)\sum_a g_a (\Lambda')-\ln(\Lambda/\mu)\sum_a g_a (\Lambda')$. In order for the $\mu$ to consistently drop out to leading order, we see that there must be a universal property   $g_a(\Lambda) \sim g_a(\Lambda') \sim g_a$.

In a similar vein,  if we compute the one loop effective action
\be
e^{i\Gamma_\text{eff}[\varphi]}=\int_{k<\Lambda} D [\delta \varphi]e^{iS_\Lambda[\varphi+\delta \varphi]}
\ee
so that $\Gamma_\text{eff}[\varphi]=S_\Lambda[\varphi]+\Delta \Gamma_\text{eff}[\varphi]$, we find that any power law divergences generated in $\Delta \Gamma$ are cancelled by the corresponding terms in the Wilson action, $S_\Lambda$.  In contrast, the couplings associated with logarithmic divergences can survive as a universal feature of the effective action, necessarily independent of $\Lambda$.

These universal features of logarithmic divergences, in contrast to power law divergences,   are strongly reminiscent of the results we saw in the previous section. When we regularised a divergent series with a smoothed cut-off, we saw how it was the finite terms  that were universal,  yielding the results obtained via analytic continuation, in contrast to the power law divergences that were regulator dependent.  How far can we push the similarities? For example, having introduced the enhanced regulators to eliminate divergences  appearing in number theory, it is tempting to ask if they can also play an interesting role in the loop divergences arising in QFT and whether this can be linked to the absence of divergences in finite theories. Indeed, the reluctance of string theory to travel deep into the UV is a direct manifestation of the smoothing or smearing effect of the string length scale as seen in a study of modular invariance of the worldsheet. This was first observed in string theory by Shapiro \citep{Shapiro72}, but a more modern treatment can be found in \citep{Polchinski98}.

\subsection{$\eta$ regularisation in QFT} \label{sec:eta}

To control  the ultra-violet divergences that plague perturbative QFT, several regularisation schemes are included in the literature.  A sharp momentum cut-off connects intuitively with Wilson's understanding of the renormalisation group \cite{Kadanoff:1966wm, Wilson:1971bg,Wilson:1971dh,Wilson:1974mb} but is known to  break translational invariance and, perhaps more importantly, the gauge invariance of the underlying theory. The standard way to regularise loop integrals whilst preserving gauge invariance is dimensional regularisation \cite{tHooft:1972tcz}. However, in analytically continuing the dimensionality of the spacetime, we run into difficulties when defining quantities unique to
 four dimensions (such as the $\gamma_5$ matrix). 
 
For infinite series, the method of smoothed asymptotics amounts to replacing an infinite sum with a weighted infinite sum 
\be
\sum_{n=0}^\infty \# \rightarrow \sum_{n=0}^\infty \eta\left(\frac{n}{N} \right) \# .
\ee
The generalisation of this to divergent loops is straightforward: working in Euclidean signature, we simply replace the loop integral with a weighted loop integral
\begin{equation}
\int \ d^4k \# \rightarrow \int \ d^4k \   \eta \left(\frac{|k|}{\Lambda}\right) \# ,
\end{equation} 
where $|k|$ is the norm of the Euclidean four-momentum and $\Lambda$ is  a cut-off scale. Of course, if $\eta$ were just a step function, this would be nothing more than a sharp momentum cut-off. However, inspired by the method of smoothed asymptotics discussed in the previous section, we shall assume that $\eta$ is a smooth regulator, corresponding to a Schwartz function that equates to unity at the origin.  We dub this $\eta$ regularisation. 

The idea of smoothing the cut-off is, of course, nothing new although in most applications, this is implemented at the level of the propagator.  For example, in his derivation of the exact renormalisation group equation,  Polchinski makes use of a modified propagator that is exponentially damped at high momenta \citep{Polchinski84}.  As a modification of the integral measure,  $\eta$ regularisation is similar to dimensional regularisation albeit with the advantage that it does not require us to analytically continue the dimension of spacetime.  However, the proposed scheme has most in common with  the smooth operator regularisation method inspired by \cite{Ball88} and developed in \cite{Oleszczuk94,Liao96} using Schwinger proper time \cite{SchwingerPT}.  Gauge invariance is also preserved in this case, as it is with dimensional regularisation. This is because the regularisation is transferred to the proper time integral, leaving the (gauge invariant) momentum integral  unchanged.  For a recent application of the smooth operator regularisation and further discussion on symmetry preservation, see \cite{Xing:2022jtt}.

Does $\eta$ regularisation also preserve gauge invariance? Generically, it seems obvious that it will not.  Just as we saw for sharp momentum cut-offs, gauge invariance is typically broken as soon as we regularise the momentum integral directly.  However, if $\eta$ regularisation has any chance of opening up a better understanding of how divergences are absent in string theory, it must, at the very least,  admit a {\it preferred} set of regulators that preserve gauge invariance.  This may seem unlikely at first glance, although as we will see in a moment, it will be possible to preserve gauge invariance and key to that are the enhanced regulators that eliminate divergences.  We will also see how this connects to regularisation of Schwinger proper time integrals.

\subsubsection{Irreducible loop integrals and eliminating divergences} \label{sec:ilis}
To develop $\eta$ regularisation in more detail, it is convenient to introduce the concept of irreducible loop integrals (ILIs) \citep{Wu03,Wu04,Wu14}. In general, $n$-fold ILIs are defined as the $n$-loop integrals for which there are no longer the overlapping factors $(k_i - k_j + p)$ in the denominator of the integrand and no factors of the scalar momentum  $k^2$ in the numerator \cite{Wu14}.  In this work we  focus on regularising ultra-violet divergences at one-loop, postponing a detailed discussion of higher loops to future work \citep{PadillaSmith23}.  It was shown in \cite{Wu03} that upon use of the Feynman parameter method, all one-loop perturbative Feynman integrals of the one-particle irreducible graphs can be evaluated as the following one-fold ILIs in Minkowski spacetime:
\begin{eqnarray}
I_{-2\alpha} (\M^2)&=& \int \frac{d^4 k}{(2 \pi)^4} \frac{1}{(k^2 +\M^2)^{2 + \alpha}}, \label{ILIs1} \\ 
I_{-2 \alpha}^{ \mu \nu} (\M^2) &=& \int \frac{d^4 k}{(2 \pi)^4}  \frac{k^{\mu}k^{\nu}}{(k^2 +\M^2)^{3+\alpha}}, \label{ILIs2} \\ 
I_{-2 \alpha}^{\mu \nu \rho \sigma} (\M^2) &=& \int \frac{d^4 k}{(2 \pi)^4}  \frac{k^{\mu}k^{\nu}k^{\rho}k^{\sigma}}{(k^2 +\M^2)^{4+\alpha}}, \label{ILIs3}
\end{eqnarray}
where the subscript $(-2\alpha)$ labels the power counting dimension (of energy-momentum) with  $\alpha = -1$ and $\alpha = 0$ corresponding  to quadratic and logarithmically divergent integrals.  The mass term  $\M^2 =\M^2(m^2_1, p_1^2, \dots) $ is a function of Feynman parameters, external momenta, $p_i$ and corresponding mass scales, $m_i$.  Note that $k^2=g_{\mu\nu} k^\mu k^\nu$ where the metric $g_{\mu\nu}$ is written with mostly positive signature. 

The ILIs of tensor type are related to scalar integrals in the usual way:
\begin{eqnarray}
I_{-2 \alpha}^{ \mu \nu} (\M^2) &=&\frac14 g^{\mu\nu} \int \frac{d^4 k}{(2 \pi)^4}  \frac{k^2}{(k^2 +\M^2)^{3+\alpha}}, \label{ILIs2scal} \\ 
I_{-2 \alpha}^{\mu \nu \rho \sigma} (\M^2) &=& \frac{1}{4!} S^{\mu\nu \rho\sigma}  \int \frac{d^4 k}{(2 \pi)^4}  \frac{k^4}{(k^2 +\M^2)^{4+\alpha}}, \label{ILIs3scal}
\end{eqnarray}
where $S^{\mu\nu\rho\sigma}=g^{\mu\nu}g^{\rho\sigma}+g^{\mu\rho}g^{\nu\sigma}+g^{\mu\sigma}g^{\nu\rho}$ is a totally symmetric tensor of rank four. 
Notice that these relations are merely  a consequence of  the rotational  symmetries of the  four dimensional space and not to be confused with the gauge consistency relations derived in \citep{Wu03,Wu04,Wu14}. As we will see later,  the latter can be accommodated by using different $\eta$s for regularising different ILIs.

When implementing the $\eta$ regularisation, we  Wick rotate the integrals to Euclidean signature $k_0 \to i k_4$ and insert a factor of $\eta(|k|/\Lambda)$ in the integrand. This yields $I_{-2 \alpha }^{\cdots }\to iJ_{-2\alpha}^{\cdots}[\eta]$, where the form of $\eta$ need not be universal for all one-fold ILIs, at least in principle. In particular, we now have that
\begin{eqnarray}
J_{-2\alpha}[\eta] (\M^2) &=& \int \frac{d^4 k}{(2 \pi)^4} \frac{1}{(k^2 +\M^2)^{2 + \alpha}}\eta\left(\frac{|k|}{\Lambda}\right), \label{JLIs1} \\
J_{-2 \alpha}^{ \mu \nu}[\eta](\M^2) &=& \frac14 g^{\mu\nu}  \int \frac{d^4 k}{(2 \pi)^4}  \frac{k^2}{(k^2 +\M^2)^{3+\alpha}}\eta\left(\frac{|k|}{\Lambda}\right)   \label{JLIs2}, \\
J_{-2 \alpha}^{\mu \nu \rho \sigma }[\eta] (\M^2) &=&  \frac{1}{4!} S^{\mu\nu \rho\sigma}  \int \frac{d^4 k}{(2 \pi)^4}  \frac{k^4}{(k^2 +\M^2)^{4+\alpha}}\eta\left(\frac{|k|}{\Lambda}\right), \label{JLIs3}
\end{eqnarray}
where the integration is over four dimensional Euclidean space in each case. Making use of  partial fractions, the regularised tensor ILIs can be written explicitly in terms of scalar counterparts  
 \begin{eqnarray}
 J_{-2 \alpha}^{ \mu \nu}[\eta](\M^2)  & =&\frac14 g^{\mu\nu}\left[J_{-2\alpha}[\eta] (\M^2)-\M^2 J_{-2(\alpha+1)}[\eta] (\M^2)\right], \label{decomp1} \\
 J_{-2 \alpha}^{\mu \nu \rho \sigma }[\eta] (\M^2)  &=&\frac{1}{4!} S^{\mu\nu \rho\sigma} \left[J_{-2\alpha}[\eta] (\M^2)-2\M^2 J_{-2(\alpha+1)}[\eta] (\M^2) +\M^4 J_{-2(\alpha+2)}[\eta] (\M^2)  \right] \qquad \quad \label{decomp2}.
 \end{eqnarray}
To understand the role of divergences, it is now sufficient to study the scalar integrals of the form \eqref{JLIs1}.  After integrating out the three-sphere, these can be written as
\be
J_{-2\alpha}[\eta] (\M^2)=\frac{1}{8 \pi^2} \int_0^\infty  dk \ \frac{k^3}{(k^2+\M^2)^{2+\alpha}}  \eta\left(\frac{|k|}{\Lambda}\right).
\ee
For $\alpha>0$ the integrals are convergent as $\Lambda \to \infty$ and one readily obtains
\be
J_{-2\alpha}[\eta] (\M^2) \sim \frac{1}{16 \pi^2 \alpha (1+\alpha)\M^{2\alpha}}. \label{ascon}
\ee
For $\alpha \leq 0$, the integrals diverge as  $\Lambda \to \infty$, where they take the following asymptotic form
\begin{eqnarray}
J_{0}[\eta] (\M^2) &\sim& \frac{1}{8 \pi^2} \left[ \ln (\Lambda/|\M|) +\gamma[\eta] -\frac12 \right], \label{as0} \\
J_{2}[\eta] (\M^2) &\sim& \frac{1}{8 \pi^2} \left[ \Lambda^2 C_{1}[\eta]-\M^2\left(  \ln (\Lambda/|\M|) +\gamma[\eta]\right) \right], \label{as2}
\end{eqnarray}
and 
\be
J_{2s+4}[\eta] (\M^2) \sim \frac{1}{8 \pi^2} \left[ \sum_{z=0}^s \binom{s}{z}  C_{2z+3}[\eta]\M^{2(s-z)}\Lambda^{2z+4} \right] \label{asn}
\ee
for any natural number $s$. These expressions are valid provided $\eta$ is a regulator: a smooth Schwartz function with $\eta(0) = 1$. Here we have also defined the  finite integral
\be \label{gammanint}
\gamma[\eta]=-\int_0^\infty dx  \  \eta'(x)  \ln x,
\ee
and recall that $C_z[\eta]=\int_0^\infty x^z \eta(x)$ is the corresponding Mellin transform for  any natural number $z$.  We immediately see the parallels with the divergent series in the previous section. Power law divergences are regulator dependent and weighted by the corresponding Mellin transform. These can always be eliminated with a judicious use of enhanced regulators. Logarithmic divergences are independent of the regulator, just as the finite terms were in the divergent series.  This universal form for the log divergences agrees with the corresponding terms obtained by other regularisation methods, such as dimensional regularisation.  Of course,  for our QFT integrals, there are also finite terms, although unlike those that appear in the divergent series, they are regulator dependent. This is a direct consequence of the logarithm and the correction to the finite term that arises when we rescale the cut-off $\Lambda$.  Indeed, for any regulator $\eta(x)$ and real number $\lambda>0$, we can define another regulator $\eta_\lambda(x)=\eta(\lambda x)$ which is equivalent to rescaling the cut-off. It then follows that $\gamma[\eta_\lambda]=\gamma[\eta]-\ln \lambda$ while $C_z[\eta_\lambda]=\lambda^{-1-z} C_z[\eta]$.  These relations will be useful when we consider the role of gauge invariance in the next section.

\subsection{Gauge invariant $\eta$ regularisation}\label{VacuumPolarisation}
To investigate how gauge invariance is affected by $\eta$ regularisation, we follow \cite{Wu03, Wu04, Wu14} and consider a general gauge theory where the gauge group has dimension $d_G$ and where $N_f$ Dirac spinors $\Psi_n$ ($n=1, \ldots, N_f$) are interacting with the Yang Mills field $A^a_\mu$ ($a=1, \ldots, d_G$). Such a theory is described by a Lagrangian
\be
\mathcal{L}=\bar \psi_n (i\gamma^\mu D_\mu-m)\psi_n -\frac14 F_{\mu\nu}^a F^{\mu\nu}_a,
\ee
where
\be
F^a_{\mu\nu}=\partial_{\mu} A_\nu^a-\partial_{\nu} A_\mu^a - gf_{abc} A^b_\mu A^c_\nu, \qquad D_\mu \psi_n=(\partial_\mu +ig T^a A^a_\mu )\psi_n,
\ee
and  $T^a$ are the generators of the gauge group whose commutator $[T^a, T^b]=i f^{abc}T^c$ defines the structure constants $f^{abc}$. A careful computation of the vacuum polarisation  for the gauge field at one-loop yields an expression of the form \cite{Wu03, Wu04, Wu14}
\be
\Pi^{ab}_{\mu\nu}(p)=\Pi^{(g) ab}_{\mu\nu}(p)+\Pi^{(f) ab}_{\mu\nu}(p),
\ee
where $p^\mu$ is the external momentum. Here   $\Pi^{(g) ab}_{\mu\nu}(p)$ are the pure Yang Mills contributions coming from  gauge field loops and ghost loops.  $\Pi^{(f) ab}_{\mu\nu}(p)$ are the contributions from fermion loops, arising from the interaction of the fermions with the gauge field.  Gauge invariance is understood in terms of the Ward identities $p^\mu\Pi^{ab}_{\mu\nu}=\Pi^{ab}_{\mu\nu}p^\nu=0$.  Requiring this to hold for any gauge theory and with any number of fermions means that Ward identities should hold separately for the gauge field and fermionic contributions
\be \label{condition}
 p^\mu\Pi^{{(g)} ab}_{\mu\nu}=\Pi^{(g)ab}_{\mu\nu}p^\nu=0, \qquad  p^\mu\Pi^{{(f)} ab}_{\mu\nu}=\Pi^{(f)ab}_{\mu\nu}p^\nu=0.
\ee
When the vacuum polarisation is computed in terms of the regularised ILIs, $I^{\cdots}_{-2\alpha}|_\text{regularised}$, these generalized Ward identities  impose strict consistency conditions  on the regularisation scheme. There are also generalised Ward identities and corresponding consistency conditions associated with three and four point functions. Altogether,  we find that for $\alpha=-1, 0, 1$ we must have (see Appendix \ref{AppendixA} and \cite{Wu03, Wu04, Wu14,Xing:2022jtt} for further details)
\begin{eqnarray}
I^{\mu\nu}_{-2\alpha}|_\text{regularised} &\sim&\frac{1}{2(\alpha+2)} g^{\mu\nu} I_{-2\alpha}|_\text{regularised}  \label{cond1}, \\
I^{\mu\nu\rho \sigma}_{-2\alpha}|_\text{regularised} &\sim &\frac{1}{4(\alpha+2)(\alpha+3)} S^{\mu\nu\rho\sigma} I_{-2\alpha}|_\text{regularised}, \label{cond2}
\end{eqnarray}
in the asymptotic limit. We now wish to apply these conditions in the context of $\eta$ regularisation. As mentioned earlier, it is important to note that, in principle, different ILIs use different regulators. At this stage, we do this in order to keep things sufficiently general.  As we will see shortly,  the Ward identities impose relations between different $\eta$s for different loop topologies, which may point to a deeper underlying structure. We will investigate the meaning of these relations in more detail in our forthcoming work \cite{PadillaSmith23}.

In the following we denote the regulators for scalar ILIs $I_{-2\alpha}$ by  $\eta_{-2\alpha}$ and the regulators for the corresponding tensor ILIs of rank two and four by  $\theta_{-2\alpha}$ and $\kappa_{-2\alpha}$ respectively. It follows that 
\begin{eqnarray}
 I_{-2\alpha}|_\text{regularised} &=& i J_{-2\alpha}[\eta_{-2\alpha} ], \label{Jeta} \\
I^{\mu\nu}_{-2\alpha}|_\text{regularised} &=&  i J^{\mu\nu}_{-2\alpha} [\theta_{-2\alpha} ], \label{Jtheta} \\
I^{\mu\nu\rho \sigma}_{-2\alpha}|_\text{regularised} &=& iJ^{\mu\nu\rho \sigma}_{-2\alpha}[\kappa_{-2\alpha} ], \label{Jkappa}
\end{eqnarray}
We now impose the conditions \eqref{cond1} and \eqref{cond2} making use of the decomposition formulae \eqref{decomp1} and \eqref{decomp2} giving
\begin{eqnarray}
\frac{\alpha}{\alpha+2} J_{-2\alpha} \left[ \tilde \theta_{-2\alpha}\right]&\sim & \M^2 J_{-2(\alpha+1)}[\theta_{-2\alpha}] \label{Jcond1}, \\
\frac{\alpha(\alpha+5)}{(\alpha+2)(\alpha+3)} J_{-2\alpha} \left[ \tilde \kappa_{-2\alpha}\right]& \sim & 2\M^2 J_{-2(\alpha+1)}[\kappa_{-2\alpha}]-\M^4 J_{-2(\alpha+2)}[\kappa_{-2\alpha}],  \label{Jcond2}
\end{eqnarray}
as $\Lambda \to \infty$  and where 
\be
\tilde \theta_{-2\alpha}=\frac{(\alpha+2)\theta_{-2\alpha}-2 \eta_{-2\alpha}}{\alpha}, \qquad \tilde \kappa_{-2\alpha}=\frac{(\alpha+2)(\alpha+3)\kappa_{-2\alpha}-6 \eta_{-2\alpha}}{\alpha(\alpha+5)}. \label{regs}
\ee
Note that for $\alpha\neq 0$ the combination of regulators appearing in \eqref{regs}  are themselves regulators,  being Schwartz and equating to unity at the origin, $\tilde \theta_{-2\alpha}(0)=\tilde \kappa_{-2\alpha}(0)=1$. It remains to plug in the  asymptotic expressions for the scalar ILIs given by \eqref{ascon}, \eqref{as0}, \eqref{as2} and \eqref{asn}. For the convergent ILIs, with $\alpha \geq 1$,  we can use the asymptotic formula \eqref{ascon} to show that the consistency relations hold automatically in the limit as $\Lambda \to \infty$, as, of course, they should.  For $\alpha=0$,  the conditions \eqref{Jcond1} and $\eqref{Jcond2}$ can be understood via the $\alpha \to 0$ limit, giving
\begin{eqnarray}
J_{0} [ \theta_{0}]- J_{0} [ \eta_{0} ]&\sim & \M^2 J_{-2}[\theta_{0}], \label{Jcond1_}\\
J_{0} [ \kappa_{0}]- J_{0} [ \eta_{0} ] & \sim & 2\M^2 J_{-2}[\kappa_{0}]-\M^4 J_{-4}[\kappa_{0}].  \label{Jcond2_}
\end{eqnarray}
Plugging in the asymptotic formulae, we see that  there are logarithmic divergences, although they cancel and the leftover finite parts yield the following constraints
\be \label{gamma0}
\gamma[\theta_0]-\gamma[\eta_0]=\frac14, \qquad \gamma[\kappa_0]-\gamma[\eta_0]=\frac{5}{12}.
\ee
For $\alpha=-1$, we get both quadratic and logarithmic divergences.  The latter cancel when  we impose the  two consistency conditions \eqref{Jcond1} and \eqref{Jcond2}, just as they did for $\alpha=0$, and we are left with
\begin{eqnarray}
\Lambda^2 C_1[\tilde \theta_{2}]+2\M^2\left[\gamma[\theta_2]-\gamma[\eta_2]-\frac14\right] &\sim& 0, \label{J2cond1} \\
\Lambda^2 C_1[\tilde \kappa_2]+\frac{3}{2} \M^2\left[\gamma[\kappa_2]-\gamma[\eta_2]-\frac{5}{12}\right]  &\sim & 0, \label{J2cond2}
\end{eqnarray}
where we recall that $\tilde \theta_{2}=2\eta_2-\theta_2$ and $\tilde \kappa_2=\frac32 \eta_2-\frac12 \kappa_2$.  Cancellation of the quadratic divergences requires that $\tilde \theta_2$ and $\tilde \kappa_2$ are enhanced regulators of order one
\be \label{C1eqn}
C_1[\tilde \theta_{2}]=C_1[\tilde \kappa_2]=0.
\ee
Here we see for the first time the connection between gauge invariance and the elimination of quadratic divergences in both divergent series and ILIs. The remaining finite parts in \eqref{J2cond1} and \eqref{J2cond2} yield constraints that are very similar to \eqref{gamma0}
\be \label{gamma2}
\gamma[\theta_2]-\gamma[\eta_2]=\frac14, \qquad \gamma[\kappa_2]-\gamma[\eta_2]=\frac{5}{12}.
\ee
At this time, we have not been able to develop any further insight into the consistency conditions arising for the finite parts, given by \eqref{gamma0} and \eqref{gamma2}. However, using the fact that $\gamma[\eta_\lambda]=\gamma[\eta]-\ln \lambda$ and $C_z[\eta_\lambda]=\lambda^{-1-z} C_z[\eta]$ for any regulator $\eta_\lambda(x)=\eta(\lambda x)$ with a rescaled cut-off, we can arrive at a rather elegant solution to the full set of consistency conditions. For $\alpha \geq -1$, this is given by
\be
\eta_{-2\alpha}(x)=\eta_{[1]}(x), \quad \theta_{-2\alpha}(x)=\eta_{[1]}(\lambda x), \quad \kappa_{-2\alpha}(x)=\eta_{[1]}(\mu x),
\ee
where $\lambda=e^{-1/4}$, $\mu=e^{-5/12}$, and $\eta_{[1]}(x)$ is any enhanced regulator of order one.

\subsubsection{Connecting to Schwinger proper time} \label{sec:SPT}
It was noted in \cite{Xing:2022jtt} that regularised Schwinger proper time integrals  provide an elegant framework for satisfying Wu's consistency relations \eqref{cond1} and \eqref{cond2}.  In particular, we write the regularised ILIs as follows
\begin{eqnarray}
I_{-2\alpha} (\M^2)|_\text{regularised}&=& \frac{i}{(\alpha+1)!} \int_0^\infty \frac{d\tau}{\tau} \ \rho(\Lambda^2 \tau)  \tau^{2 + \alpha} \int \frac{d^4 k}{(2 \pi)^4}  e^{-\tau (k^2+\M^2)}, \label{ILIs1ST} \\ 
I_{-2 \alpha}^{ \mu \nu} (\M^2) |_\text{regularised}&=& \frac{i}{(\alpha+2)!} \int_0^\infty \frac{d\tau}{\tau} \ \rho(\Lambda^2 \tau)  \tau^{3 + \alpha} \int \frac{d^4 k}{(2 \pi)^4}  k^\mu k^\nu e^{-\tau (k^2+\M^2)}, \label{ILIs2ST} \\ 
I_{-2 \alpha}^{\mu \nu \rho \sigma} (\M^2) |_\text{regularised}&=&  \frac{i}{(\alpha+3)!} \int_0^\infty \frac{d\tau}{\tau} \ \rho(\Lambda^2 \tau)  \tau^{4 + \alpha} \int \frac{d^4 k}{(2 \pi)^4}  k^{\mu}k^{\nu}k^{\rho}k^{\sigma} e^{-\tau (k^2+\M^2)},\quad
 \label{ILIs3ST}
\end{eqnarray}
where  $\rho(u) \to 0$ faster than any power as $u\to 0$ and $\rho(u) \to 1$ as $u \to \infty$. To make contact with the current work, we note that we can write these formulae in the form of equations \eqref{Jeta} to \eqref{Jkappa}, provided we generalise the regulators as follows:
\begin{eqnarray}
\eta_{-2\alpha}(x) \to \eta_{-2\alpha}(x, \M_\Lambda)&=&\frac{\int_0^\infty \frac{du}{u} \rho(u) [u(x^2+\M_\Lambda^2)]^{\alpha+2} e^{-u(x^2+M_\Lambda^2)}}{\int_0^\infty\frac{du}{u} \rho(u) [u\M_\Lambda^2]^{\alpha+2} e^{-u M_\Lambda^2}}, \\
\theta_{-2\alpha}(x) \to \theta_{-2\alpha}(x, \M_\Lambda)&=&\frac{\int_0^\infty \frac{du}{u} \rho(u) [u(x^2+\M_\Lambda^2)]^{\alpha+3} e^{-u(x^2+M_\Lambda^2)}}{\int_0^\infty \frac{du}{u} \rho(u) [u\M_\Lambda^2]^{\alpha+3} e^{-u M_\Lambda^2}}, \\
\kappa_{-2\alpha}(x) \to \kappa_{-2\alpha}(x, \M_\Lambda)&=&\frac{\int_0^\infty \frac{du}{u} \rho(u) [u(x^2+\M_\Lambda^2)]^{\alpha+4} e^{-u(x^2+M_\Lambda^2)}}{\int_0^\infty  \frac{du}{u} \rho(u) [u\M_\Lambda^2]^{\alpha+4} e^{-u M_\Lambda^2}},
\end{eqnarray} 
where $u=\Lambda^2 \tau$,  $x=|k|/\Lambda$, and $\M_\Lambda=|\M|/\Lambda$.  These do not have the precise form of the $\eta$ regulators described in this paper as the $\Lambda$ dependence is spread between  $x=|k|/\Lambda$ and $\M_\Lambda=|\M|/\Lambda$. We will consider much more general regulators in our forthcoming work \cite{PadillaSmith23}, but for now let us examine the form of these regulators in the limit where $\M_\Lambda \to 0$.   In particular, note that in this limit, the denominators can generically be written as 
\begin{eqnarray}
\int_0^\infty  \frac{du}{u} \rho(u) [u\M_\Lambda^2]^{n} e^{-u M_\Lambda^2} &=& \int_0^\infty  \frac{dy}{y} \rho\left(\frac{y}{\M_\Lambda^2}\right) y^{n} e^{-y} \\
&\sim & \int_{0^+}^\infty  \frac{dy}{y}  y^{n} e^{-y} =(n-1)!
\end{eqnarray}
It follows that as $\M_\Lambda \to 0$, the regulators fall into the class described by equation \eqref{lambdan}, with
\be \label{SPTregs}
\eta_{-2\alpha}(x, 0)=\lambda_{2+\alpha}(x), \qquad \theta_{-2\alpha}(x, 0)=\lambda_{3+\alpha}(x), \qquad \kappa_{-2\alpha}(x, 0)=\lambda_{4+\alpha}(x).
\ee
Using the expression \eqref{enhlam}, it is easy to verify that  $\tilde \theta_2=2\eta_{2}(x, 0)-\theta_{2}(x, 0)$  and  $\tilde \kappa_2=\frac32\eta_{2}(x, 0)-\frac12\kappa_{2}(x, 0)$ are enhanced regulators of order one, consistent with the gauge consistency condition \eqref{C1eqn}.  Of course,  this had to be true given the claim in \cite{Xing:2022jtt} that regularised Schwinger proper time integrals automatically satisfy Wu's consistency relations but it serves as a nice check of our formalism.

We can also compute the finite integral \eqref{gammanint} for each of the regulators  in \eqref{SPTregs}.  Indeed, from \eqref{lambdan}, we change the integration variable to $y=ux^2$ and write 
\be
\lambda_n(x)=\frac{1}{(n-1)!}\int_0^\infty \frac{dy}{y}  \rho(y/x^2)  y^{n} e^{-y},
\ee
to  then find that 
\begin{eqnarray}
\gamma[\lambda_n]&=&\frac{1}{(n-1)!}  \int_0^\infty  dx \int_0^\infty dy   \rho'(y/x^2)  \frac{y^{n}}{x^3} e^{-y} (2 \ln x) \\
&=& \frac{1}{2(n-1)!}     \int_0^\infty dy  \int_0^\infty  du   \rho'(u)  y^{n-1} e^{-y} ( \ln y-\ln u)  \\
& =& \frac{1}{2} \left[\Psi(n) - \int_0^\infty du \rho'(u) \ln u \right], 
\end{eqnarray}
where $\Psi(z)=\Gamma'(z)/\Gamma(z)$ is the digamma function and we have used the fact that  $\rho(u) \to 0$  as $u\to 0$ and $\rho(u) \to 1$ as $u \to \infty$. It immediately follows that 
\begin{eqnarray}
\gamma[\theta_{-2\alpha}(x, 0)]-\gamma[\eta_{-2\alpha}(x, 0)]&=&\frac{1}{2\alpha+4} \label{gammaST1},\\
\gamma[\kappa_{-2\alpha}(x, 0)]-\gamma[\eta_{-2\alpha}(x, 0)]&=&\frac{2 \alpha +5}{2 \left(2+\alpha \right) \left(\alpha +3\right)}. \label{gammaST2}
\end{eqnarray}
For the logarithmically divergent ILIs corresponding to $\alpha=0$, the expressions \eqref{gammaST1} and \eqref{gammaST2} satisfy the finite parts of the consistency condition, given by  \eqref{gamma0}. However, for the quadratically divergent ILIs corresponding to $\alpha=-1$, they do not agree with the corresponding consistency conditions \eqref{gamma2}. The difference is easily understood.  Although \cite{Xing:2022jtt}  has shown that  ILIs regularised using Schwinger proper time automatically satisfy Wu's consistency relations, recall that this is not exactly equivalent to $\eta$ regularisation with regulators of the form $\eta(|k|/\Lambda)$ for which we derived the consistency conditions \eqref{C1eqn},  \eqref{gamma0} and \eqref{gamma2}. The equivalence only emerges in the limit $\M_\Lambda \to 0$, suggesting that in the limit of large $\Lambda$, only the leading order constraints should align. This is indeed the case. For quadratically divergent ILIs, the leading order constraint is \eqref{C1eqn} and this holds for the derived regulators, $\eta_{2}(x, 0)$, $\theta_{2}(x, 0)$ and $\kappa_{2}(x, 0)$, whereas the subleading constraint \eqref{gamma2} does not, exactly as expected.  For logarithmically divergent ILIs,  the leading order constraint is \eqref{gamma0}, which  we have just seen to hold for the derived regulators, $\eta_{0}(x, 0)$, $\theta_{0}(x, 0)$ and $\kappa_{0}(x, 0)$, just as it should.

\section{Conclusions} \label{sec:conc}

We have explored possible connections between analytic number theory and the study of divergent series and the ultra-violet regularisation of loop integrals in perturbative quantum field theory.  On the number theory side, we have extended Tao's work on smoothed asymptotics \cite{Tao11} which offers a tantalising taste of QFT. Both exhibit regulator dependence of power law divergences, while the universal features of finite terms in Tao's study of divergent series mirror the universal features of logarithms in QFT. However, our analysis runs much deeper than this elegant analogy.  Inspired by Tao's work, we have developed a new and general method for regularising divergent  integrals in QFT which we dub $\eta$ regularisation. As a result, we have demonstrated a connection between the regularisation of divergent series and the elimination of divergences in analytic number theory and the preservation of gauge invariance at one loop in a regularised quantum field theory. Is this just a coincidence, or does it signal something deeper and a possible window into how divergences are eliminated at high energies in a complete microscopic theory?  

There is certainly much more to be learnt before we can answer this question. Indeed, it is notable that the consistency relations for preserving gauge invariance, at least in the context of $\eta$ regularisation, also include a constraint on the finite terms, as well as the divergences. We have not been able to glean a deeper understanding of the meaning of these additional constraints, beyond noting that they are satisfied in the appropriate limit by $\eta$ regulators inspired by Schwinger proper time integrals.

The link to Schwinger proper time integrals is particularly intriguing, not least because of an intuitive connection between the worldline formulation of QFT and the field theory limit of string theory.  We also note how regularised Schwinger proper time techniques have been used to improve the behaviour of functional RG equations for gauge theories at the UV cut-off \cite{deAlwis:2017ysy}. It would be interesting to see if a functional RG equation derived using enhanced $\eta$ regulators enjoyed similar properties and to explore the implications for Weinberg's asymptotic safety program.

In a forthcoming publication, we will extend our study of $\eta$ regularisation to a much broader class of regulators,  going beyond the structure inspired by Tao \cite{PadillaSmith23}. This will allow us to capture all known methods  of regularisation and more beyond. Implementing Wu's consistency relations \cite{Wu03,Wu04,Wu14} in this general setting will also be shown to yield a master equation for the regularisation scheme in the form of a differential equation. Finding solutions will be the same as finding gauge invariant regulators, which will include dimensional regularisation, as well as the enhanced $\eta$ regulators discussed here. We will also describe how $\eta$ regularisation can be understood in Lorentzian signature, as well as its extension to arbitrary loop order, consistent with unitarity, locality, and causality.

Let us say a little more about higher loops and, in particular, difficulties associated with overlapping divergences that begin at two-loop order. Key insights in this regard were developed in the symmetry preserving regularisation scheme developed by Wu \cite{Wu03}, which has much in common with our generalised approach. In a future study of how to apply $\eta$ regularisation to higher loop graphs, careful treatment of overlapping divergences and subdivergences will exploit generalisations of the ILI formalism. Moreover, at one-loop we have seen that the ILIs encode the overall UV contribution and tensor structure of Feynman integrals. Regularisation then follows from a simple redefinition of the measure in the master integrals. At two-loops and higher, this shall be seen to be similarly true albeit in a much more general way. In the repeated use of Feynman parametrisation an important step will be to ensure no Feynman parameter integration contains UV divergences and that appropriate subtractions can be made to show all overlapping divergences are harmless. Using the $\alpha,\beta,\gamma$ technology first introduced by ‘t Hooft and Veltman \cite{Hooft72}, and following the procedures set out in \cite{Wu03}, including key theorems based on the ILI formalism for factorisation, subtraction, and reduction of overlapping divergences, we will show  how $\eta$ regularisation can be extended to meet such demands. It will be seen how the treatment of overlapping Feynman integrals in generalised $\eta$ regularisation requires certain technical requirements different from those found in the loop regularisation by Wu \cite{Wu03}.

There are several ways in which we may continue to build the bridge towards string theory and the softening of amplitudes at high energies.  Given that critical string theory is a fundamentally higher dimensional theory, it would be interesting to explore the dimensional dependence of Wu's consistency relations and the implications on $\eta$ regularisation. It is also important to better understand the role played by gravity.  Is there a gravitational analogue of Wu's consistency relations that guarantees diffeomorphism invariance is preserved by the regulator? Further, how do we implement  $\eta$ regularisation in curved space? 

The work of \citep{Abel19,Abel20} could open up a path connecting $\eta$ regularisation (in the Schwinger representation) to string theory via non-local particle theories that preserve the higher dimensional properties of strings.  In this context, we are particularly interested in those regulators  that exhibit some remnant of modular invariance, such as those presented in \citep{Abel21}. We would also like to deepen this analysis and  say something about the nature of QFT and the effect of smoothing in position space.

Although our main focus is on developing QFT and building a bridge towards string theory, there is much to be explored at the level of analytic number theory. For example, it would be interesting to see if any other gauge invariant regulators identified in \cite{PadillaSmith23} have a role to play in the suppression of divergences in divergent series. Further, how do these enhanced regulators connect to analytic continuation? Tao has already provided some insight in this regard \cite{Tao11}. To develop  more understanding it would be useful to extend Tao's results to series of non-polynomial functions such as logarithms and polylogarithms, which are also expected to appear in applications to QFT. What relation, if any, can then be made to resurgence theory and trans-series? In particular, it would be interesting to see if any connection can be made between $\eta$ regularisation and some general resummation procedure that might restore some, or all, non-perturbative information. 

\begin{appendices}
\section{Vacuum polarisation and generalised ward identities}\label{AppendixA}
Following \cite{Wu03}, the contribution of $N_f$ equal mass fermion loops to the gauge boson self-energy is given by
\begin{equation}\label{Vacpolarisationfunct}
\Pi^{(f) ab}_{\mu \nu}(p) = 4 N_f   \Tr(T^a T^b) (ig)^2 \int \frac{d^4k}{(2\pi)^4} \frac{\Tr[\gamma_{\mu}(\slashed{k} + m)\gamma_{\nu}(\slashed{k} + \slashed{p} + m)]}{(k^2 + m^2 - i\epsilon)((k + p)^2 + m^2 - i\epsilon)}.
\end{equation}
After expressing the colour charge factor as $\Tr(T^a T^b) = C_2 \delta^{ab}$  for some constant $C_2$ and computing the Dirac algebra, we use Feynman parametrisation to write the remaining integrals in terms of one-fold ILIs as clearly laid out in \cite{Wu03}. Doing so gives
\begin{equation}\label{VacpolILI}
\Pi^{(f) ab}_{\mu \nu}(p) =  -4N_f g^2 C_2 \delta^{ab}  \int_0^1 dx \ \bigg[ 2 I_{2 \mu \nu }(\M) - g_{\mu \nu} I_{2} (\M) + 2x(1-x)(p^2 g_{\mu \nu} - p_{\mu}p_{\nu})I_{2}(\M) \bigg],
\end{equation}
where the mass factor $\M = m^2 -x(1-x)p^2$ in the ILIs includes a contribution from the fermion mass $m$, the external momentum, $p_\mu$ and the Feynman parameter,  $x$.  The derivation of this formula requires several comments. Here and in \cite{Wu03}, the formula is presented in terms of unregulated integrals.  However, these integrals are divergent. Strictly speaking,  the chain of manipulations can only be done for convergent integrals, which in our case would correspond to a particular form of $\eta$ regularisation \footnote{We thank Killian Moehling for pointing this out in detail.}. In future work we will explore whether or not this choice can generalise some of our conclusions. Further, one might also worry that the regulator breaks the invariance under translations of loop momenta. However, as we will explain in more detail in \cite{PadillaSmith23}, $\eta$ regulators can be rendered translation invariant by redefining the origin of momentum space.    

Taking the regularised form of  \eqref{VacpolILI}, we have that
\begin{multline}\label{VacpolILIreg}
\Pi^{(f) ab}_{\mu \nu}(p)|_\text{regularised} \propto  \int_0^1 dx \ \bigg[ 2 I_{2 \mu \nu }(\M)|_\text{regularised} - g_{\mu \nu} I_{2} (\M) |_\text{regularised} \\ + 2x(1-x)(p^2 g_{\mu \nu} - p_{\mu}p_{\nu})I_{0}(\M)|_\text{regularised} \bigg],
\end{multline}
As stated in the main text, we now require that this expression satisfies the generalised Ward identity  \eqref{condition} asymptotically, 
\begin{equation}
    p^\mu\Pi^{{(f)} ab}_{\mu\nu}=\Pi^{(f)ab}_{\mu\nu}p^\nu=0.
\end{equation}
We immediately see that the logarithmically divergent terms on the second line  of \eqref{VacpolILIreg} vanish automatically when contracted with the  external momenta and do not affect  the generalised Ward identity \eqref{condition}. However, this is not the case with the two quadratically divergent terms on the first line of \eqref{VacpolILIreg}. These must cancel asymptotically  in order to preserve gauge invraiance, giving the condition \eqref{cond1} for $\alpha=-1$, 
\begin{equation}
    I_{2}^{\mu \nu }(\M)|_\text{regularised} \sim \frac12  g^{\mu \nu} I_{2} (\M) |_\text{regularised}
\end{equation}
Note that the full set of conditions given by \eqref{cond1} and \eqref{cond2} for $\alpha=-1,0, 1$ are obtained from the remaining generalised Ward identities coming from gauge field contributions to the vacuum polarisation tensor and from higher point correlation functions. 
\end{appendices}

\section*{Acknowledgements}
We are grateful to John Barrett, Pete Millington, Killian Moehling, Lubos Motl, Benjamin Muntz, Terence Tao and Paul Saffin for useful comments and discussions. RGCS was supported by a Bell Burnell Studentship and AP  by STFC consolidated grant number ST/T000732/1.  For the purpose of open access, the authors have applied a CC BY public copyright licence to any Author Accepted Manuscript version arising. No new data were created during this study.

\bibliographystyle{apsrev4-2}
\bibliography{QFTpapers}
\end{document}